\documentclass[aps,prl,twocolumn,showpacs,preprintnumbers,
superscriptaddress,nobalancelastpage,nofootinbib]{revtex4}

\usepackage{graphicx}
\usepackage{amssymb,amsmath}

\usepackage{epstopdf}
\begin{document}



\title{Upper Bound on the Dark Matter Total Annihilation Cross Section}

\author{John F. Beacom}
\affiliation{Department of Physics, Ohio State University,
Columbus, Ohio 43210}
\affiliation{Department of Astronomy, Ohio State University,
Columbus, Ohio 43210}
\affiliation{Center for Cosmology and Astro-Particle Physics,
Ohio State University, Columbus, Ohio 43210}

\author{Nicole F. Bell}
\affiliation{California Institute of Technology,
Pasadena, California 91125}
\affiliation{School of Physics, The University of Melbourne, 
Victoria 3010, Australia}

\author{Gregory D. Mack} 
\affiliation{Department of Physics, Ohio State University,
Columbus, Ohio 43210}
\affiliation{Center for Cosmology and Astro-Particle Physics,
Ohio State University, Columbus, Ohio 43210}

\date{4 August 2006; revised 17 September 2007}

\begin{abstract}
  We consider dark matter annihilation into Standard Model particles
  and show that the least detectable final states, namely neutrinos,
  define an upper bound on the total cross section.  Calculating the
  cosmic diffuse neutrino signal, and comparing it to the measured
  terrestrial atmospheric neutrino background, we derive a strong and
  general bound.  This can be evaded if the annihilation
  products are dominantly new and truly invisible particles.
  Our bound is
  much stronger than the unitarity bound at the most interesting
  masses, shows that dark matter halos cannot be significantly
  modified by annihilations, and can be improved by a factor of
  10--100 with existing neutrino experiments.
\end{abstract}

\pacs{95.35.+d, 98.62.Gq, 98.70.Vc, 95.85.Ry}


\maketitle


%
The self-annihilation cross section is a fundamental property of dark
matter.  For thermal relics, it sets the dark matter mass
density, $\Omega_{DM} \sim 0.3$, and in these and
more general non-thermal scenarios, also
the annihilation rate in gravitationally-collapsed dark matter
halos today~\cite{BHS}.  How large can the dark matter annihilation cross section be?
There are two general constraints that bound the rate of dark matter
{\it disappearance}.  (Throughout, we mean the cross section
averaged over the halo velocity distribution, i.e., $\langle \sigma_A
v \rangle$, where $v_{rms} \sim 10^{-3}c$.) 

The first is the unitarity bound, developed for the early universe
case by Griest and Kamionkowski~\cite{Griest}, and for the
late-universe halo case by Hui~\cite{Hui}.  In the plane of $\langle
\sigma_A v \rangle$ and dark matter mass $m_\chi$, this allows only
the region below a line $\langle \sigma_A v \rangle \sim 1/m_\chi^2$
(this will be made more precise below).  The second is provided by the
model of Kaplinghat, Knox, and Turner (KKT)~\cite{KKT}, in which
significant dark matter annihilation is invoked to resolve a
conflict between predicted (sharp cusps) and observed (flat cores)
halo profiles.  Since this tension may have been
relaxed~\cite{BHS}, we reinterpret this type of model as an upper
bound, allowing only the region below a line $\langle \sigma_A v
\rangle \sim m_\chi$.  That the KKT model requires $\langle \sigma_A v
\rangle$ values $\gtrsim 10^7$ times larger than the natural scale for
a thermal relic highlights the weakness of the unitarity bound in the
interesting GeV range.  However, there have been no other strong and
general bounds to improve upon these.

\begin{figure}[t]
\includegraphics[width=3.25in]{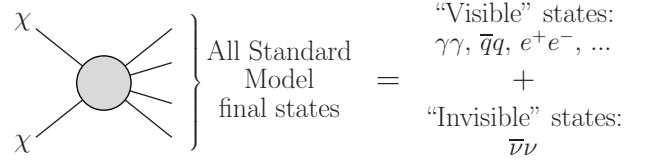}
\caption{Annihilation of dark matter into SM final states.   Since all final
states except neutrinos produce gamma rays (see text), we can bound
the total cross section from the neutrino signal limit, i.e., assuming
Br(``Invisible'') $\simeq 100\%$.}
\label{fig:diagram}
\end{figure}

While these bound the {\it disappearance} rate of dark matter, they
say nothing about the {\it appearance} rate of annihilation products,
instead assuming that they can be made undetectable.  To
evade astrophysical limits, the branching ratios to specific final
states can be adjusted in model-dependent ways.  However, a
model-independent fact is that the branching ratios for all final
states must sum to 100\%.  A reasonable assumption
is that these final states are Standard Model (SM) particles.
We show that the most difficult SM final state to detect is
neutrinos; but that surprisingly strong flux limits can be simply
derived from recent high-statistics data; and that we may interpret
these as bounding all SM final states, and hence the dark matter
{\it total} annihilation cross section.  See Fig.~\ref{fig:diagram}.

If dark matter is not its own antiparticle and if there is a large
particle-antiparticle asymmetry, then annihilation could be
prohibited, making all bounds inapplicable or irrelevant.
Our bound can be evaded if the final states
are dominantly new and truly invisible non-SM particles, in
which case all dark matter annihilation searches will be
more challenging; we quantify an upper bound on the branching
ratio to SM final states below.

{\bf Probing Dark Matter Disappearance.---}
For dark matter that is a thermal relic, the cross section required to
ensure $\Omega_{DM} \sim 0.3$ is $\langle \sigma_A v \rangle \sim 3
\times 10^{-26} \,\, \text{cm}^3 \text{ s}^{-1}$~\cite{BHS}.  KKT discussed
several models in which the dark matter is not a thermal relic, e.g.,
it might have acquired mass only in the late universe, or have been
produced through the late decays of heavier particles~\cite{KKT}.  As
emphasized in Refs.~\cite{KKT, Hui}, it is interesting to directly
ask how large the
annihilation cross section could be in halos today, irrespective of
possible early-universe constraints.

Unitarity sets a general upper bound on $\langle \sigma_A v \rangle$,
and can only be evaded in certain unusual cases~\cite{Griest, Hui, Kusenko}.
In the low-velocity limit where
the cross section is assumed to be s-wave dominated, $\langle \sigma_A
v \rangle \leq 4 \pi / m_\chi^2 v$, or
\begin{equation}
\langle \sigma_A v \rangle 
\leq 1.5 \times 10^{-13} \frac{\text{ cm}^3}{\text{s}}
\left[\frac{\text{GeV}}{m_\chi}\right]^2 
\left[\frac{300 \text{ km/s}}{v_{rms}}\right]\,.
\end{equation}
In the KKT model, the required cross section to sufficiently distort the dark
matter profiles of galaxies is
\begin{equation}
\langle \sigma_A v \rangle_{\rm KKT}
\simeq 3 \times 10^{-19} \frac{\text{ cm}^3}{\text{s}} \left[\frac{m_\chi}{\text{GeV}}\right]\,.
\end{equation}
(Similar effects are attained via elastic interactions~\cite{Spergel}.
A large self-annihilation cross section implies a large elastic
self-scattering cross section, but not vice-versa~\cite{Hui}.)
Hui~\cite{Hui} showed that unitarity restricts the KKT model to relatively
small masses; for $v_{rms} = 300$ km/s, $m_\chi \lesssim$ 80 GeV.
There have been no other model-independent methods to constrain the
KKT model.  We argue next that dark matter disappearance must be
accompanied by the appearance of something, and the bound on the
weakest final state bounds all of them.
The appearance rate bounds $\langle \sigma_A v \rangle$ directly,
independent of which partial waves dominate $\sigma_A$, i.e., its
$v$-dependence.


{\bf Revealing Neutrino Appearance.---}
We assume that annihilation proceeds to SM particles, and
express the cross section in terms of branching ratios
to ``visible'' and ``invisible'' final states, such as gamma rays and
neutrinos, respectively, as in Fig.~\ref{fig:diagram}.
If the branching ratio to a specific final state were known, then a
bound on that appearance rate would yield a bound
on the total cross section, inversely proportional to this branching
ratio.  However, the branching ratios are model-dependent, and
any specific one can be made very small, making that bound on
$\langle \sigma_A v \rangle$ very weak, e.g., for $m_\chi = 1$ GeV,
KKT require $Br(\gamma) \lesssim 10^{-10}$ to allow their total cross
section~\cite{KKT}.
Note that gamma-ray data constrain only the product
$\langle \sigma_A v \rangle \, Br(\gamma)$, 
and the bounds vary with $m_\chi$.

KKT~\cite{KKT} and Hui~\cite{Hui} assume invisible but
unspecified final states.  It is clear that most SM final states produce
gamma rays.  Quarks and gluons hadronize, producing pions, where
$\pi^0 \rightarrow \gamma \gamma$; the decays of weak bosons and tau
leptons also produce $\pi^0$.
The stable final state $e^+ e^-$ is not
invisible, since it produces gamma rays either through
electromagnetic radiative corrections~\cite{RC} or energy loss
processes~\cite{Eloss}; the final state $\mu^+ \mu^-$
produces $e^+ e^-$ by its decays.
Thus the only possible ``invisible" SM final states are neutrinos.

Of final states
with neutrinos, we focus on $\bar{\nu} \nu$.  Similar bounds
could be derived for $\bar{\nu} \bar{\nu} \nu \nu$, but we assume that
these are suppressed and/or that the Rube Goldberg-ish Feynman
diagrams required would contain charged particles, and hence gamma
rays through (model-dependent) radiative corrections.  Due to
electroweak bremsstrahlung, final-state neutrinos are inevitably
accompanied by weak bosons and hence gamma rays, primarily with
$E_\gamma \simeq m_\pi/2$; however, these gamma-ray constraints on
$\langle \sigma_A v \rangle$ are weaker than or comparable to what
we obtain directly with neutrinos~\cite{Kachelriess}.

To derive our bound on the {\it total}
annihilation cross section, we assume $Br(\bar{\nu} \nu) \simeq 100\%$.
This is {\it not} an assumption about realistic outcomes, but it is the
right way to derive the {\it most conservative} upper bound for SM
final states.
Why is this a bound on the total cross section, and not just on the
partial cross section to neutrinos?
Suppose that $Br(\bar{\nu} \nu)$ were reduced enough that 
the $1/Br(\bar{\nu} \nu)$ correction for an impure final state
was necessary; at our factor-two precision, this occurs when
another SM final state has a comparable branching ratio.
For the total cross section set by the neutrino
bound, any other pure final state would be more strongly
constrained, thus making this cross section disallowed for
all final states in the SM.
Therefore, while setting this bound using neutrinos can be too
conservative, it can never overreach.


{\bf Cosmic Diffuse Neutrinos: Signal.---}
The most direct approach to bound the $\chi \chi \rightarrow \bar{\nu}
\nu$ cross section is to use the cosmic diffuse neutrino flux from
dark matter annihilations in all halos in the universe as the signal.
Since this is isotropic and time-independent, it is challenging to
detect above the background caused by the atmospheric
neutrino flux.
A complementary approach uses the Milky Way signals, which
have somewhat different uncertainties on the predictions and
data~\cite{DMhalo}.  The data to test the diffuse signal are
available in the full energy range now, but this is not yet true
for all the directional signals.  While the latter will likely be
stronger eventually, going beyond our rough estimates will
require proper experimental analyses.

The cosmic diffuse signal from $\chi \chi \rightarrow
\bar{\nu} \nu$ annihilations depends on the radial density
profile of each dark matter halo, the halo mass function (the relative
weighting of halos of different masses), and how those halos evolve
with redshift.  We follow the calculations of Ullio {\it et
  al.}~\cite{UllioPRD, UllioPRL}; see also~\cite{others}.  The
signal spectrum is
\begin{equation}
\frac{d\Phi_\nu}{dE} = \frac{\langle \sigma_A v\rangle}{2}
\frac{c}{4\pi H_0} \frac{\Omega_{DM}^2\rho_{\rm crit}^2}{m_\chi^2}
\int_0^{z_{\text{up}}} \! dz \frac{\Delta^2(z)}{h(z)}
\frac{dN_\nu(E')}{dE'},
\label{ullioeq}
\end{equation}
where the 1/2 is for assuming $\chi$ is its own antiparticle, $H_0=100~h
\, \text{km} \, \text{s}^{-1} \, \text{Mpc}^{-1} $ is the Hubble
parameter, and $\Omega_{DM}$ is the dark matter density in units of
the critical density $\rho_{\rm crit}$.  We assume a flat universe,
with $\Omega_{DM}=0.3$, $\Omega_\Lambda=0.7$, $h=0.7$, and $h(z)=
[(1+z)^3 \Omega_{DM} + \Omega_\Lambda]^{1/2}$.

The factor $\Delta^2(z)$ accounts for the increase in density due to
the clustering of dark matter in halos, defined so that $\Delta^2 = 1$
corresponds to all dark matter being at its average density in the
universe today.  The concentration of halos and thus the value of
$\Delta^2$ evolves with redshift.  (Note that we have absorbed a
factor of $(1+z)^3$ into the definition of $\Delta^2$, as in
Ref.~\cite{UllioPRL}.)  However, to collect most of the signal, we
only need neutrino energies near $m_\chi$, and hence will be
sensitive only to modest redshifts where it is accurate to take
$\Delta^2(z) \simeq \Delta^2(0)$~\cite{UllioPRD}.
Note that only $\Delta^2$ matters, and not its individual factors.
The value of $\Delta^2$ does depend on the halo profile chosen.
We adopt $\Delta^2 = 2 \times 10^5$ for Navarro, Frenk, and
White (NFW) halos with moderate assumptions about the halo
mass distribution.  For cuspier Moore profiles, $\Delta^2$ could
be $\simeq 10$ times larger, while for flatter Kravtsov profiles,
it could be $\simeq 2$ times smaller; see the discussions in
Ref.~\cite{DMhalo}.

\begin{figure}[t]
\includegraphics[width=3in]{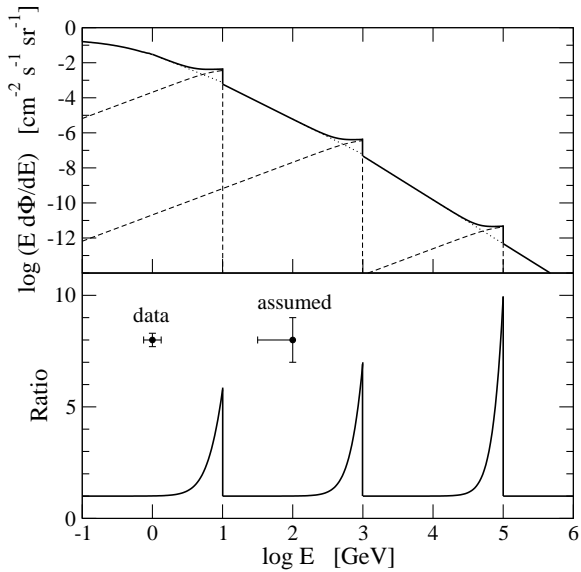}
\caption{{\bf Upper:} Diffuse $\bar{\nu} \nu$ annihilation signal for
  $m_\chi = 10, 10^3,$ and $10^5$ GeV, added to the atmospheric
  background, both as ($\bar{\nu}_\mu + \nu_\mu$) and versus neutrino
  energy.  As noted, the signals are most accurate for $E_\nu \gtrsim
  m_\chi/3$.  {\bf Lower:} Ratio of this sum and background.  The
  $\langle \sigma_A v \rangle$ values at each example $m_\chi$ are
  chosen to be detectable by our conservative criteria; the data and
  assumed uncertainty scales are also indicated.}
\label{fig:spectrumplot}
\end{figure}

For $\chi \chi \rightarrow \bar{\nu} \nu$, the source spectrum
$dN_\nu/dE'$ is a delta function; neutrinos produced with energy $E'$
are redshifted to the observed energy $E = E' / (1+ z)$, i.e.,
\begin{equation}
\label{branch}
\frac{dN_\nu(E')}{dE'} = \frac{2}{3} \delta(m_\chi-E') 
= \frac{2}{3E}\delta\left[z-\left(\frac{m_\chi}{E}-1\right)\right]\,,
\end{equation}
where we have accounted for 2 neutrinos per annihilation, equally
divided among 3 flavors.  (Note that $\nu_\mu$ has a
large fraction in every neutrino mass eigenstate, so any initial
mix of mass or flavor eigenstates would be close to this.)  
In Fig.~\ref{fig:spectrumplot}, we show example dark matter
signals compared to the atmospheric neutrino background.

Using the neutrino signal, we can also derive
a constraint from the relativistic energy
density.  Requiring $\Omega_{\text{rad}} < 0.2$ at low
redshift~\cite{Zentner} leads to $\langle \sigma_A v \rangle <
10^{-17} \times (m_\chi / \text{GeV}) \text{ cm}^3/\text{s}$.  While
this bound applies to {\it any} light final state
including non-SM particles such as purely sterile neutrinos,
it is weak, and would require even greater halo modifications
than the KKT model.


{\bf Cosmic Diffuse Neutrinos: Backgrounds.---}
How large of a neutrino signal is allowed by present data?  As shown
in Fig.~\ref{fig:spectrumplot}, the signal spectrum is sharply peaked.
To be insensitive to the spectrum shape, i.e., the redshift evolution,
we define the signal as integrated over a bin of width $\Delta
\log_{10}{E_\nu} = 0.5$, just below $E_\nu = m_\chi$, (i.e., we
consider $z \lesssim 2$, so we can ignore the tails at low energy.)
To be detectable, we require that the
signal be 100\% as large as the angle-averaged atmospheric neutrino
($\nu_\mu + \bar{\nu}_\mu$) background, integrated in the same way.
Signal and background are both somewhat smeared from
received neutrino energy to detected energy,
but well within this bin.  This conservative approach allows us to
simply derive the flux and annihilation cross section constraints over
the very wide mass range 0.1--$10^5$ GeV.  Our model predicts equal
fluxes of $(\nu_e + \bar{\nu}_e)$, $(\nu_\mu + \bar{\nu}_\mu)$, and
$(\nu_\tau + \bar{\nu}_\tau)$, any of which can be used to derive
bounds.

The atmospheric neutrino $(\nu_\mu + \bar{\nu}_\mu)$ spectra as a
function of neutrino energy have been derived from data from the the
Fr\'ejus (0.25--$10^4$ GeV, in 9 bins)~\cite{Frejus} and AMANDA
($1.3 \times 10^3$--$3.0 \times 10^5$ GeV, in 10 bins)~\cite{AMANDA}
detectors.  Neutrino attenuation in Earth will only be significant
above $10^5$ GeV.  The agreement with theoretical predictions against
upward fluctuations in the data is very good, well below the 100\%
uncertainty that we adopted.  These spectra were derived from
neutrino-induced muon data by a regularized unfolding technique,
which might miss a narrow signal.

We thus considered the data in more detail, finding that for
$E_\nu = 0.1$--$10^4$ GeV, such a signal is definitely excluded,
especially using both
the $(\nu_\mu + \bar{\nu}_\mu)$ and $(\nu_e + \bar{\nu}_e)$ signals.
The most useful data are from the Super-Kamiokande detector.  In
Ref.~\cite{SK}, visible-energy spectra for each of $e$-like and
$\mu$-like events from 0.1--100 GeV are given in 4 log-spaced bins per
decade. The agreement with predictions including neutrino oscillations
is excellent; the moderate exceptions in some of the highest-energy
bins are explainable~\cite{SK}.  Neutrinos with $E_\nu \sim 10$--$10^3$
GeV are probed by the count rates (no spectra) of upward throughgoing
muons~\cite{SK}, and similarly for $E_\nu \sim 10^2$--$10^4$ GeV and upward
showering muons~\cite{SK-HE}; both are also in excellent agreement
with predictions.

Dedicated analyses of the measured data could improve the signal
sensitivity by a factor 10--100, depending on the energy range.  First,
using the sharp
feature in the spectrum at $m_\chi$; Fig.~\ref{fig:spectrumplot} shows
that while the signal is comparable to the background when integrated
over an energy bin, in the endpoint region it is much larger.
Second, the uncertainties below 10 GeV are actually below
10\%, and apply to narrower bins in energy than we assumed~\cite{SK}.
Third, by 10 GeV, the $(\nu_e + \bar{\nu}_e)$ to $(\nu_\mu +
\bar{\nu}_\mu)$ background flux ratio is 1/3 and rapidly
falling~\cite{AtmNu}; in addition, the $(\nu_e + \bar{\nu}_e)$ flux is
strongly peaked at the horizon~\cite{AtmNu}, while the dark matter
signal is isotropic.  Fourth, detailed analyses of Super-Kamiokande
and AMANDA upward throughgoing and upward showering muon
data should be more sensitive than the simple count rates we used.

\begin{figure}[t]
\includegraphics[width=3in]{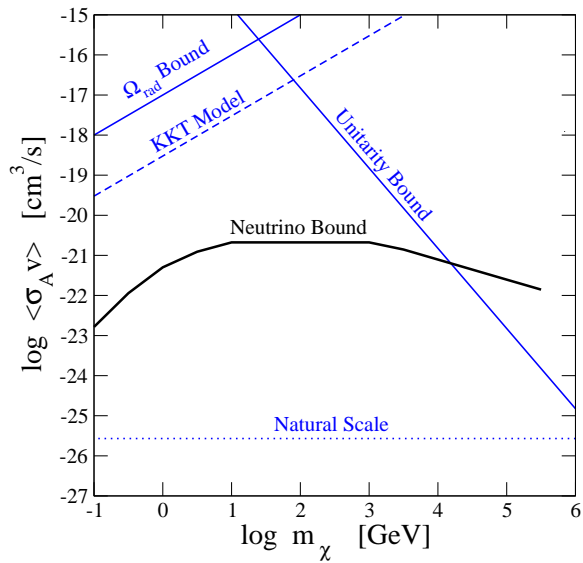}
\caption{Upper bounds on the dark matter total annihilation cross section in
galaxy halos as a function of the dark matter mass, calculated as discussed
in the text.}
\label{fig:sigmavplot}
\end{figure}

{\bf Conclusions.---}
We have shown that the dark matter total annihilation cross section
in the late universe,
i.e., the dark matter {\it disappearance} rate, can be directly and
generally bounded by
the least detectable SM states, i.e., the neutrino {\it appearance} rate.
This can be simply and robustly constrained by comparing the diffuse
signal from all dark matter halos to the terrestrial
atmospheric neutrino background.  Our bound on $\langle \sigma_A v
\rangle$ is shown in Fig.~\ref{fig:sigmavplot}.
Over a large range in $m_\chi$, 
it is much stronger than the
unitarity bound of Hui~\cite{Hui}.  It strongly rules out the proposal of
Kaplinghat, Knox, and Turner~\cite{KKT} to modify dark matter halos by
annihilation.  Our bound can be evaded with truly invisible non-SM
final states.  For a cross section
above our bound, its ratio to our bound yields an upper limit
on the branching ratio to SM final states
required to invoke that large of a cross section.

Annihilation flattens halo cusps to cores of density
$\rho_A \sim m_\chi/(\langle \sigma_A v \rangle H_0^{-1})$~\cite{KKT}.
Our bound implies that for all $m_\chi \agt 0.1$ GeV, this density
is $\rho_A \agt 5 \times 10^3 \text{ GeV/cm}^3$, which only occurs
at radii $\alt 1 $ pc in the Milky Way for an NFW profile.
Annihilation should thus have minimal effects on galactic halos.

Detailed analyses by the Super-Kamiokande and AMANDA Collaborations
should be able to improve our bound by a
factor 10--100 over the whole mass range.  Halo substructure or
mini-spikes around intermediate-mass black holes could increase the
signal by orders of magnitude~\cite{others, MoreSignal}.  The
sensitivity could thus become close to the natural scale for thermal
relics, making it a new tool for testing even standard scenarios.


\medskip
We are grateful to G.~Bertone for very helpful discussions
and collaboration on an early stage of this project.  We thank
S.~Ando, S.~Desai, M.~Kachelriess, M.~Kaplinghat, E.~Komatsu,
R.~Scherrer, and H.~Y\"uksel
for helpful discussions.
JFB was supported by NSF CAREER Grant PHY-0547102;
NFB by a Sherman Fairchild fellowship at Caltech
and by the Melbourne Research Grants Scheme;
and GDM by DOE Grant DE-FG02-91ER40690.
We also thank CCAPP and OSU for support.


\end{document}